\input ./style/arxiv-general.cfg
\documentclass[aos,MSNbibl,nameyear,dvips]{arximspdf}
\makeatletter
   \@ifpackageloaded{graphicx}{}{\usepackage{graphicx}}
\makeatother
\usepackage{dcolumn}
\usepackage{url,breakurl}

\doi{10.1214/14-AOS1303}
\volume{43}
\issue{4}
\pubyear{2015}
\firstpage{1471}
\lastpage{1497}
\docsubty{FLA}

\makeatletter
\newcolumntype{d}[1]{D{,}{}{#1}}
\newcommand{\rrvert}{\vert}
\newcommand{\llvert}{\vert}
\newtheorem{lemma}{Lemma}
\newtheorem{theorem}{Theorem}
\newproclaim{remark}{Remark}
\newtheorem{proposition}{Proposition}
\newcommand{\cor}{\operatorname{cor}}
\def\T{\mathrm{T}}
\newcommand{\diff}{\mathrm{d}}
\newcommand{\bSigma}{\bolds{\Sigma}}
\newcommand{\bX}{\mathbf{X}}
\newcommand{\bS}{\mathbf{S}}
\newcommand{\bbeta}{\bolds{\beta}}
\newcommand{\balpha}{\bolds{\alpha}}
\newcommand{\bD}{\mathbf{D}}
\newcommand{\bG}{\mathbf{G}}
\newcommand{\bI}{\mathbf{I}}
\newcommand{\bT}{\mathbf{T}}
\newcommand{\bJ}{\mathbf{J}}
\makeatother

\begin{document}
\begin{frontmatter}

\title{The fused Kolmogorov filter: A nonparametric model-free
screening method}
\runtitle{The fused Kolmogorov filter}

\begin{aug}
\author[A]{\fnms{Qing}~\snm{Mai}\thanksref{T2}\ead[label=e1]{mai@stat.fsu.edu}}
\and
\author[B]{\fnms{Hui}~\snm{Zou}\corref{}\thanksref{T3}\ead[label=e2]{zouxx019@umn.edu}}
\runauthor{Q. Mai and H. Zou}
\affiliation{Florida State University and University of Minnesota}
\address[A]{Department of Statistics\\
Florida State University\\
Tallahassee, Florida 32306\\
USA\\
\printead{e1}}

\address[B]{School of Statistics\\
University of Minnesota\\
Minneapolis, Minnesota 55455\\
USA\\
\printead{e2}}
\end{aug}
\thankstext{T2}{Supported by the FYAP grant from Florida State University.}
\thankstext{T3}{Supported an NSF grant and an ONR grant.}

%
\received{\smonth{10} \syear{2014}}

%
\begin{abstract}
A new model-free screening method called the fused Kolmogorov filter is proposed
for high-dimensional data analysis. This new method is fully
nonparametric and
can work with many types of covariates and response variables,
including continuous,
discrete and categorical variables. We apply the fused Kolmogorov
filter to deal
with variable screening problems emerging from a wide range of
applications, such as multiclass
classification, nonparametric regression and Poisson regression, among others.
It is shown that the fused Kolmogorov filter enjoys the sure screening property
under weak regularity conditions that are much milder than those
required for many
existing nonparametric screening methods.
In particular, the fused Kolmogorov filter can still be powerful when covariates
are strongly dependent on each other. We further demonstrate the
superior performance
of the fused Kolmogorov filter over existing screening methods by
simulations and real data examples.
\end{abstract}

%
\end{frontmatter}

\section{Introduction}\label{sec1}
Consider a statistical problem with a response variable $Y$ and
covariates $\bX=(X_1,\ldots,X_p)^T\in\mathbb{R}^p$. When $p$ is
very large, a popular assumption is the sparsity assumption that only a
small subset of variables are actually responsible for modeling $Y$. To
be specific, following \citet{distcor}, define
\[
\bD=\bigl\{j\dvtx  F(y\mid\bX)\mbox{ functionally depends on $X_j$ for some $y$}\bigr\},
\]
where $F(y\mid\bX)$ is the conditional cumulative probability
function of $Y$. Then the sparsity assumption states that $\llvert \bD
\rrvert\ll p$.

Variable selection aims to discover $\bD$ exactly. Variable screening
is less ambitious in that it only aims to discover a majority of $\bD
^c$. In other words, a good variable screening method tries to find a
subset $\bS$ such that $\bD\subset\bS$, which is referred to as the
sure screening property [\citet{FL2008}] in the literature.
Consistent variable selection is a very challenging task. It requires
sophisticated estimation techniques, strong model assumptions and often
advanced computing algorithms [\citet
{Tibs96a,FL2001,FL2009,Zhang}]. Because variable screening deals with
a much less ambitious goal, it is possible that sure screening could be
achieved by using some simple (both conceptually and computationally)
method. This idea was first successfully demonstrated in \citet{FL2008}
where marginal correlation screening is shown to lead to sure screening
results in high-dimensional linear regression under certain regularity
conditions. Since the sure independence screening paper by \citet
{FL2008}, variable screening has received a lot of attention in the
literature and many variable screening techniques, both parametric and
nonparametric, have been proposed and studied in recent years
[\citet
{FF2008,FSW2009,FS2010,FFS11,distcor,rankcor,modelfree,Kfilter,CTW13,HWH13}].
Variable screening is naturally appealing to practitioners, because if
sure screening is achieved before doing a thorough analysis, the
analysis part would become much easier with the screening subset. At
least, the computational cost can be greatly reduced.

The main message in \citet{FL2008} is that although we should not do
variable selection based on marginal correlations alone, marginal
correlations can be used to filter out many noise variables and keep
all important variables.
Many new screening methods have been proposed with the aim of improving
the marginal correlation screening method. \citet{FS2010} propose
a screening method based on the marginal maximum likelihood for
generalized linear models. \citet{CTW13} propose using marginal
empirical likelihood ratios to rank variables and demonstrate their
good performance. The nonparametric independence screening (NIS)
[\citet{FFS11}] starts with a generalized additive model for
modeling the regression response variable $Y$. For each variable~$X_j$,
NIS uses nonparametric smoothing, for example, B-spline regression, to
obtain $\hat m_j=\arg\min_{m_j} \Vert Y-m_j(X_j)\Vert_n^2$. NIS then
selects the variables with large $\Vert\hat m_{j}(X_j)\Vert_n^2$.
Compared to marginal correlation learning, NIS is more robust because
it captures nonlinear dependence between $Y$ and $X_j$. The
quantile--adaptive screening (QA) [\citet{HWH13}] further improves
the robustness of NIS by allowing heteroscedasticity in the model.
Under such models, QA minimizes the check function instead of the
squared error loss function to identify the important predictors. \citet
{rankcor} propose using Kendall tau correlation to replace the usual
Pearson correlation in marginal correlation screening so that the
resulting screening method is more robust and can be useful under a
semiparametric single-index model with a monotone link function. The
distance correlation screening (DCS) [\citet{distcor}] is a
model-free screening method that uses the distance correlation to
replace Pearson correlation in marginal correlation screening. The
distance correlation [\citet{SRB07}] between two random variables
is zero if and only if they are independent. The Kolmogorov filter
[\citet{Kfilter}] is a fully nonparametric robust screening
method. It deals with binary classification problems and uses the
Kolmogorov--Smirnov test statistic to screen covariates. The Kolmogorov
filter has several unique, nice properties. First, it significantly
outperforms other existing screening methods for binary classification
problems. Second, it works with all types of covariates and is
invariant under univariate monotone transformations of the covariates.
Third, it can have the sure screening property even when the covariates
are strongly dependent on each other. This result is very promising
because it was commonly believed before \citet{Kfilter} that marginal
screening methods tend to work well if and only if the noise variables
are weakly correlated with the important variables.

\citet{FL2008} suggest an iterative screening and model fitting
procedure to deal with the strong correlation issue in model-based
screening methods. Although this idea has been empirically demonstrated
[\citet{FL2008,FS2010,FFS11,HWH13}], its theoretical
justification still remains unknown. Furthermore, its theoretical
justification heavily depends on model assumptions and hence may not be
very robust. It is now clear that variable screening can be separated
from the model fitting part. Both DCS [\citet{distcor}] and the
Kolmogorov filter [\citet{Kfilter}] have demonstrated that sure
screening can be achieved without resorting to a particular form of
model for the data. Moreover, we advocate the use of model-free
screening methods in practice. The reasons are twofold. First, the
model-free screening results are much more robust in the sense that the
sure screening property can hold under much weaker conditions. The
second reason is related to the choice of the statistical analysis tool
in the modeling stage. Note that after the screening we have a
low-dimensional dataset, and one may want to apply modern nonparametric
learning methods such as boosting and random forest for further
analysis [\citet{HTF2009}]. Yet model-based screening methods
typically eliminate such choices because one has to stick with the
model used in the first stage. For example, if we apply marginal
correlation screening or marginal maximum likelihood screening, we have
to use a linear regression model or generalized linear model in the
second stage, although we can do penalized model fitting by using a
penalty such as lasso [\citet{Tibs96a}] or SCAD [\citet
{FL2001}]. If the underlying model for the data is highly nonlinear,
then boosting or random forest is expected to be a better choice than
linear models.

Our goal here is to develop a new fully nonparametric model-free
variable screening method that can provide a unified solution to
variable screening problems emerging from a wide variety of
applications such as binary classification, multiclass classification,
regression and Poisson regression, among others. The new method should
also work with discrete, categorical or continuous covariates.
Moreover, it is desirable to have the new method be invariant under
univariate monotone transformations of response variable or covariates
or both, because variable transformation models have wide applications
in practice. Imagine that a variable transformation model is determined
to be the best fit in the second modeling stage, we do wish to see that
variable screening results should remain unchanged if we would repeat
the screening procedure by working with the transformed variables. DCS
and the Kolmogorov filter are the two existing, fully nonparametric,
model-free screening methods in the literature. Neither of them
completely meets our expectations. DCS does not have the invariance
property under monotone variable transformation, and its sure screening
property heavily depends on a distribution assumption on covariates
that they should have sub-exponential tails. In many applications, the
covariates are heavy-tailed, and DCS may not be ideal in such cases.
The limitation of the Kolmogorov filter is obvious as well: it is
designed for binary classification problems and is inapplicable when
the response variable can take more than two values.

To this end, we propose the fused Kolmogorov filter and study its
theoretical and numerical properties. As the name suggests, the fused
Kolmogorov filter is built upon two main ideas, the Kolmogorov--Smirnov
test statistic, as used in \citet{Kfilter}, and fusion.
When the response variable is binary, the fused Kolmogorov filter is
exactly the Kolmogorov filter proposed in \citet{Kfilter}, and fusion
is not needed. The fusion part becomes critically important when the
response variable is continuous. We introduce two levels of fusion. In
the first level, we slice the response variables into multiple slices,
compute a Kolmogorov--Smirnov test statistic for each pair of slices
and then take the supreme of all pairwise Kolmogorov--Smirnov test
statistics. To make the method insensitive to the slicing scheme, we
conduct the second level of fusion, where we repeat the first level for
different ways of slicing and then take the sum of their outcomes as
the final screening statistic, which we call the fused Kolmogorov
statistic. The second level of fusion is important when the response
variable is continuous or ordinal. The fused Kolmogorov filter ranks
each covariate by its fused Kolmogorov statistic and screens out those
covariates at the bottom of the rank list. By definition, the fused
Kolmogorov filter is intuitively appealing, computationally convenient
and automatically has the invariance property under monotone variable
transformation.

The rest of the paper is organized as follows. The methodological
details of the fused Kolmogorov filter are given in Section~\ref{sec2}. In
Section~\ref{sec3} we establish the sure screening property of the fused
Kolmogorov filter under weak regularity conditions. We discuss these
regularity conditions and find that they can hold, even when important
variables and noise variables are strongly dependent. This promising
result suggests that marginal variable screening could be more useful
than we expected.
Sections~\ref{sec4} and \ref{sec5} contain simulated and real data examples. Technical
proofs are presented in the \hyperref[append]{Appendix}.

\section{Method}\label{sec2}
\subsection{Motivation}\label{sec2.1}
To see why the Kolmogorov--Smirnov statistic is very useful for
variable screening, let us first revisit the binary Kolmogorov filter.
When the response variable is binary, say $Y=1,2$, a variable $X$ is
independent of $Y$ if and only if the conditional distributions of $X$
given $Y=1$ or $Y=2$ are identical. Motivated by this simple fact,
\citet{Kfilter} propose using
\[
\label{K1} K_j=\sup_x\bigl\llvert
F_j(x\mid Y=1)-F_j(x\mid Y=2)\bigr\rrvert
\]
to measure the dependence between $X_j$ and $Y$, where $F_j$ denotes
the generic cumulative distribution function (CDF) for $X_j$. Given the
observed data, an empirical version of $K_j$ is defined as
\[
\label{K2} \hat K_j=\sup_x\bigl\llvert
\hat F_j(x\mid Y=1)-\hat F_j(x\mid Y=2)\bigr\rrvert,
\]
where $\hat F_j$ denotes the generic empirical CDF. \citet{Kfilter}
demonstrate the strong theoretical and numerical performance of the
binary Kolmogorov filter.

Given the success of the binary Kolmogorov filter, it is natural to ask
what its counterpart is for a continuous response variable or
a general discrete variable (like counts data in Poisson regression).
First, it seems straightforward to consider
%
%
\begin{equation}
\label{Klimit} K^*_j=\sup_{y_1,y_2}\sup
_x\bigl\llvert F_j(x\mid Y=y_1)-F_j(x
\mid Y=y_2)\bigr\rrvert
\end{equation}
because $K_j^*=0$ if and only if $X_j$ is independent of $Y$. Thus
$K^*_j$ is a natural generalization of $K_j$. In order to use $K^*_j$,
we must have an empirical version of $K^*_j$. This step is trivial for
the binary response case, but it is much more difficult when $Y$ takes
infinite values because it requires the knowledge of $F_j(x\mid y)$ for
all possible values $y$. On the other hand, we can find an
approximation of $K_j^*$ by slicing the response.
Define a partition
\[
\bG=\Biggl\{[a_l,a_{l+1})\dvtx  a_l<a_{l+1},
l=0,\ldots,G-1\mbox{ and } \bigcup_{j=1}^{G-1}
[a_l,a_{l+1})\setminus\{a_0\}=\mathbb{R}
\Biggr\},
\]
where $a_0=-\infty$ and $a_G=\infty$. Note that the interval
$(a_0,a_1)$ is open, but we abuse the notation a little by writing the
intervals $[a_l,a_{l+1})$ for all $l$. Each $[a_l,a_{l+1})$ is called a
slice. We then define a random variable $H\in\{1,\ldots,G\}$ such
that $H=l+1$ if and only if $Y$ is in the $l$th slice. In particular,
if $Y$ is discrete as in a multiclass problem, that is, $Y=1,\ldots,
G$, we can set $H=Y$. Now let
\[
\label{KG} K^{\bG}_j=\max_{l,m}\sup
_x\bigl\llvert F_j(x\mid H=l)-F_j(x
\mid H=m)\bigr\rrvert,
\]
where $F_j(x\mid H=l)=\Pr(X_j\le x\mid H=l)$.

The idea of slicing is very natural. First, If $Y$ is binary, $K^\bG
_j$ and $K_j$ are the same. If $Y$ is multiclass, the slicing breaks
the multiclass problem into pairwise binary problems. This strategy has
been proven successful as a method for generalizing a binary classifier
to its multiclass counterpart [\citet{HT98}]. Yet $K^{\bG}_j$
can be still be computed when $Y$ is a count that takes infinite
discrete values, such as in the Poisson regression model. When $Y$ is
continuous, slicing is widely used in the field of sufficient dimension
reduction [\citet{Li91,SAVE}] to infer about the conditional
means and/or variances of predictors. However, these sufficient
dimension reduction methods generally deal with problems with large
sample sizes compared to the dimension. To the best of our knowledge,
this paper is the first to utilize slicing for variable screening for
large $p$ and small $n$ problems.

It is obvious that $X_j$ is independent of $Y$ if and only if $K^\bG
_j=0$ when $Y$ takes finite values and each possible value forms a
slice. In what follows, we assume that $Y$ is continuous, as it is the
more challenging case. The following lemma shows that $K^\bG_j$ sheds
light on the dependence between $Y$ and $X_j$ as well when $Y$ is continuous.

%
\begin{lemma}\label{Kindependent}
\textup{(a)} $X_j$ is independent of $Y$ if and only if $K^{\bG}_j=0$
for all possible choices of $\bG$.

\textup{(b)} Assume that $X_j$ is not independent of $Y$ and for any
fixed $y\in\mathbb{R}$, $\Pr(Y\le y\mid X_j=x)$ is not a constant in
$x$; then $K^\bG_j\ne0$ for any $\bG$.

\textup{(c)} Assume that $F_j(x\mid y)$ is continuous in $y$. If $\max
_{l=1,\ldots, G}\Pr(H=l)\rightarrow0$ as $G\rightarrow\infty$, then
$
K^\bG_j\rightarrow K^*_j$ as $G\rightarrow\infty$, where
$K_j^*$ is defined in (\ref{Klimit}).
Therefore, for $X_j$ not independent of $Y$, $K^\bG_j>0$ for
sufficiently large $G$.
\end{lemma}

Although we initially proposed $K^\bG_j$ as a surrogate of $K^*_j$ and
Lemma \ref{Kindependent} part (c) indicates this as well, it turns out that $K^\bG_j$
could be a better measure for variable screening than $K^*_j$. To see
this interesting point, we present the following lemma.

%
\begin{lemma}\label{Knormal} 
If $(X_j,Y)$ has a bivariate Gaussian copula distribution such that,
after transformation via two monotone functions $g_1,g_2$,
$(g_1(X_j),g_2(Y))$ is jointly normal with correlations $\rho
_j=\operatorname{Cor}(g_1(X_j),g_2(Y))$ and $g_1(X_j),g_2(Y)$
are margimally standard normal. Then we have the following two conclusions:
\begin{longlist}[(a)]
\item[(a)] $K^*_j=1$ if $\rho_j\ne0$ and $K^*_j=0$ otherwise.\vspace*{1pt}
\item[(b)] Suppose Y is sliced at $\frac{l}{G}$'th quantile of $Y$
for $l=1,\ldots, G-1$. Then $K^\bG_j$ can be expressed as
\[
K^\bG_j=G\int_{-\infty}^{\Phi^{-1}(1/G)}
\biggl(2\Phi\biggl(\frac
{-\llvert \rho_j\rrvert y}{\sqrt{1-\rho_j^2}}\biggr)-1 \biggr)\frac
{e^{-y^2/2}}{\sqrt
{2\pi}}\,\diff y,
\]
where $\Phi$ is the CDF for the standard normal distribution.
Consequently, for any $G$, $K^\bG_j$ is a strictly increasing function
in $\llvert \rho_j\rrvert$.
\end{longlist}
\end{lemma}

With Lemma~\ref{Knormal} in mind, we revisit the variable screening
problem under a high-dimensional linear regression model as examined in
[\citet{FL2008}]. For simplicity, assume that the model is
\[
Y=X_1+X_2+Z %
\]
and
\[
X_j=a X_1+Z_j,\qquad j \ge3,
\]
where $X_1,X_2,Z,Z_j$ are independent $N(0,1)$ variables. Then we have
\begin{eqnarray*}
\operatorname{Cor}(X_1,Y)&=&\operatorname{Cor}(X_2,Y)=
\frac{1}{\sqrt{3}},
\\
\operatorname{Cor}(X_j,Y)&=&\frac{a}{\sqrt{3(1+a^2)}}\qquad\mbox{for }j=3,\ldots,
p.
\end{eqnarray*}
So this is a perfect case for using the marginal correlation screening
of \citet{FL2008}. By Lemma~\ref{Knormal} we have the following results:
\begin{eqnarray*}
K^*_j&=&1, \qquad j=1,2,3,\ldots,
\\
K^\bG_1&=& K^\bG_2>K^\bG_j,
\qquad j=3,\ldots. %
\end{eqnarray*}
Thus $K^*_j$ cannot separate $(X_j, j\ge3) $ from $X_1,X_2$ no matter
how small $a$ is. On the other hand, $K^\bG_j$ works perfectly in this
example, just like the marginal correlations. Of course, $K^\bG_j$ in
general works much better than $\operatorname{Cor}(X_j,Y)$, which will be
clearly demonstrated in the later sections.

\subsection{The fused Kolmogorov filter}\label{sec2.2}
In this subsection we show how to use $K^\bG_j$ for variable screening
based on a random sample $(\bX^i,Y^i)_{i=1}^n$. We first need to
estimate $K^\bG_j$ accurately for all $p$ variables. Given a partition
$\bG$, we estimate $K^\bG_j$ by
\[
\hat K^\bG_j=\max_{(l,m)}\sup
_y\bigl\llvert\hat F_j(x\mid
H_j=l)-\hat F_j(x\mid H_j=m)\bigr\rrvert,
\]
where
\[
\hat F_j(x\mid H=l) =\frac{1}{n_l}\sum
_{H^i=l} \mathrm{1}\bigl(X_j^i\le x
\bigr),
\]
and $n_l$ is the sample size within the $l$th slice, and $H^i=l$ if
$Y^i$ is in the $l$th slice.

If $Y$ is a multi-level categorical variable, then the partition is
simply done according to $Y$'s value. When $Y$ has infinitely many
possible values, the partition/slicing scheme can be important. With
finite sample size, it is important to have enough sample sizes within
each slice to control the estimation variance. As mentioned in the
\hyperref[sec1]{Introduction}, the idea of slicing response variable has been used by
researchers in sufficient dimension reduction. Early researchers proved
that the sliced inverse regression (SIR) can be consistent even when
there are only two observations in each slice [\citet
{Li91,HC92}], which implies that SIR is reasonably insensitive to the
slicing scheme. Yet \citet{ZN95} later observed that, even though SIR
can be consistent for all slicing schemes with the same number of
observations in each slice, there is a loss of efficiency when there
are too many slices. Based on our experience, the choice of slices does
not affect variable screening results very much. However, significant
improvement can be achieved by fusion.
Suppose that we have $N$ different partitions, $\bG_i$ for $i=1,\ldots, N$, where each partition $\bG_i$ contains $G_i$ intervals. Then we let
\[
\label{fusedK} \hat K_j=\sum_{i=1}^N
\hat K^{\bG_i}_j.
\]
By doing so, we combine the information from all $\bG_i$. This fusion
step is motivated by \citet{CZ13}, who showed that in sufficient
dimension reduction, combining several slicing schemes works better
than the usual practice relying on a single slicing scheme. As shown in
Section~\ref{sec4}, fusion does yield variable screening results that are
superior to using a single slicing scheme.

We suggest an intuitive uniform slicing to partition data into $G$
slices. If $Y$ is categorical with levels $1,\ldots,G$, or $Y$ is
discrete with finite possible values $1,\ldots, G$, we set $H=Y$. If
$Y$ is discrete and can take infinite values as in a Poisson regression
model, we set $H=Y+1$ if $Y<G-1$ and $H=G$ if $Y\ge G-1$.
For the case where $Y$ is continuous, we let the partition $\bG$
contain the intervals bounded by the $\frac{l}{G}$th sample quantiles
of $Y$ for $l=0,\ldots,G$. From now on, we always write $\hat
K_j(G)=\hat K_j^{\bG}$ when $\bG$ is a uniform partition with $G$ slices.
By fusion,\vspace*{1pt} we consider multiple uniform slicing $\bG_i, 1 \le i \le N$
where $\bG_i$ has $G_i$ many slices. In practice, we suggest choosing
$G_i\le\lceil\log{n}\rceil$ for all $i$ so that there is a decent
sample size within each slice for all slicing schemes. This is
important because the fused Kolmogorov filter is a fully nonparametric
method and sample size plays a central role in nonparametric
statistics. Then the final fused Kolmogorov filter statistic is
%
%
\begin{equation}
\label{fkused} \hat K_j=\sum_{i=1}^N
\hat K^{\bG_i}_j,
\end{equation}
and the fused Kolmogorov filter screening set is defined as
%
%
\begin{equation}
\label{setD} \hat\bD=\bigl\{j\dvtx \mbox{$\hat K_j$ is among the
$d_n$'th largest}\bigr\}.
\end{equation}

\section{Theory}\label{sec3}

In this section we establish the sure screening property of the fused
Kolmogorov filter.

\subsection{Main theorem}\label{sec3.1}
We first introduce a concept called the oracle fused Kolmogorov filter.
If we know the distribution of $Y$, then we can use an oracle uniform
slicing such that the partition $\bG_i$ contains the intervals bounded
by the $\frac{l}{G_i}$th theoretical quantiles of $Y$ for $l=0,\ldots,G_i$. For this special slicing, write $K_j^{(o)}(G_i)=K^{\bG_i}_j$
and $K^{(o)}_j=\sum_i K_j^{(o)}(G_i)$. Then we can obtain a screening
set as
$\hat\bD(\mbox{oracle})=\{j\dvtx  \mbox{$\hat K^{(o)}_j$ is among the
$d_n$'th largest}\}$, where $d_n$ is a predefined positive integer.
Throughout this section, $C$ denotes a generic positive constant.

To show the sure screening property of the fused Kolmogorov filter, we
consider the following two regularity conditions:

\subsubsection*{Regularity conditions}
\begin{longlist}[(C2)]
\item[(C1)] There exists a set $\bS$ such that $\bD\subset\bS$ and
\[
\Delta_{\bS}=\min_i\Bigl(\min_{j\in\bS}
K_j^{(o)}(G_i)-\max_{j\notin
\bS}
K_j^{(o)}(G_i)\Bigr)>0.
\]
\item[(C2)] Let $G_{\min}=\min_i\{G_i\}$. Then for any $b_1,b_2$
such that $\Pr(Y\in[b_1,b_2))\le2/G_{\min}$, we have
%
%
\begin{equation}
\bigl\llvert F_j(x\mid y_1)-F_j(x\mid
y_2)\bigr\rrvert\le\frac{\Delta_{\bS}}{8}
\end{equation}
for all $x$, $j$ and $y_1,y_2\in[b_1,b_2)$.
\end{longlist}

%
\begin{theorem}\label{KH1}
Assume conditions \textup{(C1)}~and~\textup{(C2)}. Define
\[
\eta=CNp\bigl(\log^2{n}\bigr)\exp\biggl(-C\frac{n\Delta_\bS^2}{\log
{n}}
\biggr)+CN\bigl(\log^2{n}\bigr)\exp\biggl(-C\frac{n}{\log^2{n}}\biggr).
\]

If $G_i \le\lceil\log{n}\rceil$ for all $i$ and $d_n \ge\llvert
S\rrvert$, we
have the following conclusions:
\begin{longlist}[(2)]
\item[(1)] For the oracle fused Kolmogorov filter, we have
%
%
\begin{equation}
\label{KOsure} \Pr\bigl(\bD\subset\hat\bD(\mathrm{oracle})\bigr)\ge
1-\eta.
\end{equation}
Therefore, the oracle fused Kolmogorov filter enjoys the sure screening
property with a probability tending to one if $\Delta_{\bS}\gg\sqrt
{\frac{\log{n}\cdot\log{(pN\log{n})}}{n}}$.\vspace*{1pt}
\item[(2)] For the fused Kolmogorov filter defined in (\ref{fkused})
and its screening set defined in (\ref{setD}), we have
%
%
\begin{equation}
\label{Ksure} \Pr(\bD\subset\hat\bD)\ge1-\eta.
\end{equation}
Therefore, the fused Kolmogorov filter enjoys the sure screening
property with a probability tending to one if
%
%
\begin{equation}
\label{Th1eq2} \Delta_{\bS}\gg\sqrt{\frac{\log{n}\log{(pN\log{n})}}{n}}.
\end{equation}
\end{longlist}
\end{theorem}

%
\begin{remark}\label{re1}
By comparing (\ref{KOsure}) and (\ref{Ksure}), we see that the
fused Kolmogorov filter can handle the same order of dimensions as the
oracle fused Kolmogorov filter. Therefore, slicing at the sample
quantiles results in a method that is as powerful as one utilizing
oracle information about the theoretical quantiles. Also, Theorem~\ref
{KH1} sheds light on the choice of $\bG_i$. The minimum number of
slices was 3 in \citet{CZ13}. Then Theorem~\ref{KH1} requires that $G_i\le
\lceil\log{n} \rceil$, with each $\bG_i$ containing $G_i$ intervals
bounded by sample quantiles. Therefore, in practice, we suggest setting
$G_i=3, \ldots, \lceil\log{n} \rceil$, with each $\bG_i$
containing $G_i$ intervals bounded by sample quantiles.
\end{remark}

%
\begin{remark}\label{re2}
One could obtain a limit on the dimension for the fused Kolmogorov
filter from Theorem~\ref{KH1}. Suppose we choose the slicing scheme recommended
in Remark~\ref{re1}. It follows that $N\le\log{n}$. Then if there exists
$0<\kappa<1$ such that $\Delta_{\bS}\gg n^{-\kappa}$, (\ref
{Th1eq2}) reduces to
\[
\label{pub} \log{p}\ll n^{\xi},
\]
for any $\xi\in(0,1-2\kappa)$. Note that this restriction on $p$ is
the same as that for SIS; see Conditions 1~and~3 in \citet{FL2008}.
Therefore, the fused Kolmogorov filter can handle the same order of
dimensions as SIS without imposing any parametric assumptions.
\end{remark}

%
\begin{remark}
Theorem~\ref{KH1} shows that the fused Kolmogorov filter enjoys the
sure screening property with a probability tending to one as long as we
choose a reasonably large $d_n$. One interesting fact is that (\ref
{Ksure}) does not involve $d_n$ explicitly. It holds as long as
$d_n\ge\llvert \bS\rrvert$. This insensitivity to $d_n$ leads to tremendous
practical convenience, because we can always use a reasonably large
$d_n$ to guarantee a high probability of enjoying the sure screening
property. In particular, when performing variable selection, one often
assumes that the number of important variables is less than $n$. For
example, lasso can only produce up to $n$ nonzero coefficients.
Therefore, when we apply the fused Kolmogorov filter, we can use
$d_n=a\lceil\frac{n}{\log{n}}\rceil$ where $a$ is some constant. A
more conservative choice could be $d_n=n$.
\end{remark}

%
\begin{remark}
With the regularity conditions (C1)--(C2), the sure screening property
results from the fact that $\hat K_j$ are close to $K^{(o)}_j$, which
is a consequence of the Dvoretzky--Kiefer--Wolfowitz inequality. In the
following subsection, we further discuss the implications of the two
regularity conditions.
\end{remark}

\subsection{Comments on the regularity conditions}\label{sec3.2}

The conditions for Theorem~\ref{KH1} are very mild. First, note that,
in contrast to DCS [\citet{distcor}], we make no assumption on
the distribution of $\bX$. Therefore, the fused Kolmogorov filter is
expected to be more powerful than DCS when the predictors are
heavy-tailed. Moreover, we do not assume any form of the dependence of
$Y$ on $\bX$. So the fused Kolmogorov filter will be more flexible
than NIS and QA. The only two conditions we require are conditions (C1)~and~(C2).

We first comment on condition (C2). This condition is slightly stronger
than requiring $F_j(x\mid y)$ to be continuous in $y$, as in Conclusion
(c) of Lemma~\ref{Kindependent}. Such a condition guarantees that the
sample quantiles of $Y$ are close enough to the population quantiles of
$Y$. Obviously, this result is expected for many distributions of~$Y$.
A conseqence is that the actual slicing used in practice is very close
(asymptotically) to the oracle slicing such that $\hat K_j$'s
accurately approximate $\hat K^{(o)}_j$'s.

In order to establish the sure screening property, a nontrivial
condition is needed. For example, the partial orthogonality condition,
that is, $\bX_{\bD}\perp\bX_{\bD^C}$ [\citet
{HTM2008,FS2010}], has been considered in the literature. Clearly, the
theory is more interesting when $\bX_{\bD}$ and $\bX_{\bD^C}$ are
dependent. In our theory,
condition (C1) is the core condition which is used to guarantee that
jointly important predictors, that is, the predictors belonging to the
set $\bD$, should also be marginally important, which is more or less
assumed in the theory for existing marginal screening methods in the
literature. In the context of binary classification, it has been shown
that the sure screening property of the Kolmogorov filter can be
established even when $\bX_{\bD}$ and $\bX_{\bD^C}$ are strongly
correlated [\citet{Kfilter}]. This phenomenon can be generalized
to the multiclass classification rather directly, whose derivation is
omitted here for the sake of space. In what follows we focus on the
case where $Y$ is continuous to show that condition (C1) can still be
true even when $\bX_{\bD}$ and $\bX_{\bD^C}$ are strongly
correlated, and hence the sure screening property can hold with high
probability. We highlight this interesting point by
considering the following variable-transformation linear normal model:
%
%
\begin{equation}
\label{semi} T_y(Y)=\bT(\bX)^\T\bbeta+\varepsilon,
\end{equation}
where $\bT=(T_1,\ldots, T_p)$ and $T_y,T_1,\ldots,T_p$ are strictly
monotone univariate transformations. It is also assumed that $\bT(\bX
)\sim N(0,\bSigma)$ with $\Sigma_{jj}=1$ for $j=1,\ldots,p$, and
$\varepsilon\sim N(0,\sigma^2)$ is independent of $\bX$. Note that
$(T_y,\bT)$ are unknown, and we do not assume any parametric forms for
them. Therefore, (\ref{semi}) is a very flexible semiparametric
regression model. The main idea in model (\ref{semi}) is that after
whitening each variable in the dataset we could fit a linear regression
model. This interesting model has close connections to many
transformation models in the literature; for example, see \citet
{BF85,HS97,rankcor}.

%
\begin{lemma}\label{correlated}
Consider the model in (\ref{semi}). Without loss of generality, assume
that $\bbeta=(\bbeta_\bD,0)$. Define $\balpha=\bSigma\bbeta$.
Then for any set of $\bG_i,i=1,\ldots, N$, we have:
\begin{longlist}[(2)]
\item[(1)] Condition \textup{(C1)} is true if and only if there exists $\bS$
such that\break $\min_{j\in\bS}\llvert \alpha_j\rrvert>\max_{j\notin\bS
}\llvert \alpha_j\rrvert$.
\item[(2)] If $\bSigma$ is blockwise diagonal, that is, $\sigma
_{ij}=0$ if $i\in\bD,j\notin\bD$, then $\Delta_\bD>0$ if and only
if $\min_{j\in\bD} \llvert \alpha_j\rrvert>0$.
\item[(3)] Suppose $\Sigma_{ij}=\rho^{\llvert i-j\rrvert}$. If $\min
_{j\in\bD
}\llvert \alpha_j\rrvert>0$ and we let
\[
\bS=\biggl\{1,\ldots,d+\biggl\lceil\frac{\log{(\min_{j\in\bD}\llvert
\alpha
_j\rrvert/\llvert \alpha_d\rrvert)}}{\log{\llvert \rho\rrvert
}}\biggr\rceil\biggr\},
\]
then $\Delta_\bS>0$.
\item[(4)] Suppose\vspace*{2pt} $\Sigma_{ij}=\rho$ and $\Sigma_{jj}=1$. Define
$\bS=\{j\dvtx \alpha_j\ne0\}$. Then $\Delta_\bS>0$. Moreover, $\bD
\subset\bS$ if and only if $\mathrm{1}^\T\bbeta=0$.
\item[(5)] Suppose $\Sigma_{ij}=\rho$ and $\Sigma_{jj}=1$. Then
$\Delta_{\bD}>0$ if $\rho>0$ and $\beta_j$ has the same sign for
all $j\in\bD$.
\end{longlist}
\end{lemma}

In the following we discuss the implications of Lemma~\ref{correlated}.

%
\begin{remark}
In\vspace*{1pt} part (3) where the covariance has an autoregressive structure, to
ensure the sure screening property, we need $d_n\ge d+\break \lceil\frac
{\log{(\min_{j\in\bD}\llvert \alpha_j\rrvert/\llvert \alpha_d\rrvert
)}}{\log{\llvert \rho
\rrvert}}\rceil$. It follows that
\[
\llvert\rho\rrvert\le\exp\biggl(\frac{\log{(\min_{j\in\bD}\llvert
\alpha_j\rrvert/\llvert \alpha
_d\rrvert)}}{d_n-d}\biggr).
\]
With\vspace*{1pt} $d_n=\lceil\frac{n}{\log{n}}\rceil$, the upper bound of
$\llvert \rho\rrvert$ tends to 1. Therefore, there is little
restriction on $\rho
$. In parts (4)~and~(5) where $\bSigma$ has the compound symmetry
structure, $\rho$
can be arbitrary as well.
\end{remark}

%
\begin{remark}
A direct calculation shows that in the fused Kolmogorov filter, $K_j$
is monotone in $\alpha_j$, while the joint importance $X_j$ is
measured by $\beta_j$. Part (2) of Lemma~\ref{correlated}
corresponds to the partial orthogonality condition under which the
important variables and noise variables are independent, so this is an
expected result. Somewhat surprisingly, parts (3)--(5) of Lemma~\ref
{correlated} show that even when the predictors are highly correlated,
condition (C1) still holds. Then by Theorem~\ref{KH1}, the fused Kolmogorov
filter will enjoy the sure screening property with high probability.
\end{remark}

%
\begin{remark}
Let us consider the normal linear model where we further assume
$T_y(Y)=Y$ and $T_j(X_j)=X_j$, that is, $Y=\bX^\T\bbeta+\varepsilon$,
where $\bX\sim N(0,\bSigma)$. Lemma~\ref{correlated} can be applied
to marginal correlation screening (SIS) and distance correlation
screening (DCS). However, the fused Kolmogorov filter is more flexible
than SIS, DCS and many other screening methods because it is invariant
under monotone transformations. Many existing screening methods, except
rank correlation screening [\citet{rankcor}], do not have this
nice invariance property. As a result, when the true mode is a
transformation normal linear model, SIS and DCS can perform poorly,
while the fused Kolmogorov filter's performance remains the same,
regardless of the transformations. We will clearly demonstrate this
point in the simulation study in Section~\ref{sec4}.
\end{remark}

\section{Simulations}\label{sec4}
\subsection{Simulation design}\label{sec4.1}
In this section, we compare the fused Kolmogorov filter with existing
screening methods on simulated datasets. In all the models, we set
$n=200,p=5000$. We consider the fused Kolmogorov filter based on
$K_j(G_i)$ for $G_i=3,\ldots,6$, because $\lceil\log{n}\rceil=6$.
When the response is continuous, we slice $Y$ at $\frac{l}{G_i}$th
sample quantiles for $l=1,\ldots,G_i-1$. We\vspace*{2pt} further include six other
successful screening methods in the literature for comparison, marginal
correlation screening (SIS) [\citet{FL2008}], nonparametric
independence screening (NIS) [\citet{FFS11}], distance
correlation screening (DCS) [\citet{distcor}], rank correlation
screening (RCS) [\citet{rankcor}], empirical likelihood screening
(ELS) [\citet{CTW13}] and the quantile--adaptive screening (QA)
[\citet{HWH13}]. In all the models, we use SIS to denote the
linear screening method. For example, if the response is continuous,
SIS is the original marginal correlation screening. For the generalized
linear model we use SIS to denote the marginal maximum likelihood
estimator (MMLE) [\citet{FS2010}]. When $Y$ is a multi-level
categorical variable, SIS fits $p$ multinomial models with the $\tt R$
package $\tt nnet$ [\citet{nnet}] and selects the predictors with
the largest deviances. With a little abuse of notation, we refer to all
these methods as SIS when it is clear from the context.

Following \citet{HWH13}, we consider $\alpha=0.5,0.75$ for QA.
We use the implementation of NIS and QA at \url
{http://users.stat.umn.edu/\textasciitilde wangx346/research/example1b.txt}.
The distance correlation is computed by the $\tt R$ package $\tt
energy$. For ELS, we use the implementation of ELS by the authors of
\citet{CTW13}.
As in \citet{FL2008}, we report the minimum number of predictors needed
to keep all the useful predictors. The results are based on 500 replicates.
We consider the following six models in this simulation study:

\begin{longlist}
\item[\textit{Model} 1.]
$T_y(Y)=\bT(\bX)^\T\bbeta+\varepsilon$, where $\bbeta
=2.8\times(1,-1,0_{p-2})$, $\bT(\bX)\sim N(0,\bSigma)$ with
$\bSigma=\mathrm{CS}(0.7)$, $\varepsilon\sim N(0,1)$ is independent of
$\bX$. We consider three sets of $(T_y, \bT)$:
\begin{longlist}[(a)]
\item[(a)] $T_y(Y)=Y, T_j(X_j)=X_j$; 

\item[(b)] $T_y(Y)=Y, T_j(X_j)=X_j^{1/9}$; 

\item[(c)] $T_y(Y)=Y^{1/9},T_j(X_j)=X_j$.
\end{longlist}
Models 1(a), 1(b) and 1(c) are examples of model (\ref{semi}) with a
compound symmetry correlation matrix of which the correlation
coefficient is $0.7$.
\end{longlist}

\begin{longlist}
\item[\textit{Model} 2.] $Y=\bT(\bX)^\T\bbeta+\varepsilon$, where $\bbeta
=0.8\times(\mathrm{1}_{10},0_{p-10})$. $\bT(\bX)\sim N(0,\bSigma)$
with $\bSigma=\mathrm{AR}(0.7)$. Again, we consider three sets of
$(T_y, \bT)$:
\begin{longlist}[(a)]
\item[(a)] $T_y(Y)=Y, T_j(X_j)=X_j$; 

\item[(b)] $T_y(Y)=Y, T_j(X_j)=\frac{1}{2}\log{X_j}$; 

\item[(c)] $T_y(Y)=\log(Y),T_j(X_j)=X_j$.
\end{longlist}
Models 2(a), 2(b) and 2(c) are examples of model (\ref{semi}) with an
autoregressive correlation matrix of which the autoregressive
correlation coefficient is $0.7$.
\end{longlist}

\begin{longlist}
\item[\textit{Model} 3 (\textit{Single index regression model}).]
$Y=(X_1+X_2+1)^3+\varepsilon$, where $X_j$'s follow the Cauchy
distribution independently and $\varepsilon\sim N(0,1)$ is independent of~$\bX$.
\end{longlist}

\begin{longlist}
\item[\textit{Model} 4 (\textit{Additive model}).]
$Y=4X_1+2\tan(\pi
X_2/2)+5X_3^2+\varepsilon$, where $X_j$'s follow $\operatorname{Unif}(0,1)$
independently and $\varepsilon\sim N(0,1)$ is independent of $\bX$.
\end{longlist}

\begin{longlist}
\item[\textit{Model} 5 (\textit{Heteroskedastic regression model}).]
$Y=2(X_1+0.8X_2+0.6X_3+0.4X_4+0.2X_5)+\exp
(X_{20}+X_{21}+X_{22})\varepsilon$, where $\varepsilon\sim N(0,1)$, and
$\bX\sim N(0,\bSigma)$ with $\bSigma=\mathrm{AR}(0.8)$. This model
is adapted from \citet{HWH13}. In \citet{HWH13}, they report the
minimum number of predictors to keep the first five predictors for QA
with $\alpha=0.5$ because QA with $\alpha=0.5$ can only detect the
predictors affecting the median. However, it is difficult to use such
information for other methods. Therefore, we report the minimum number
of predictors we need to keep all the eight important predictors for QA
with $\alpha=0.5$ too, so that it is fair to other methods.
\end{longlist}

\begin{longlist}
\item[\textit{Model} 6 (\textit{Poisson regression model}).]
$Y\sim\operatorname{Poisson}(\mu)$, where $\mu=\break \exp(\bX^\T\bbeta)$,
$\bbeta=(0.8,-0.8,\mathrm{0}_{p-2})$, $X_j\sim t_2$ independently.
The counterpart for SIS for this model is the marginal maximum
likelihood estimator (MMLE) [\citet{FS2010}]. Note that the
predictors are heavy-tailed in this model, and $Y$ may consequently
have extreme outliers. Therefore, to resolve computational issues, we
delete an observation whenever $Y>1000$ in MMLE. In addition, we
consider the Kolmogorov filter and DCS on this model because all other
methods are inapplicable to such datasets. Now, for the Kolmogorov
filter, we set $H=Y$ if $Y<2$; otherwise, $H=3$.
\end{longlist}

\begin{longlist}
\item[\textit{Model} 7 (\textit{Multiclass classification model}).]
$Y=1,\ldots, 5$. For each $g$, if $Y=g$, $X_{2(g-1)+1}$ and $X_{2g}$
independently follow $0.5 N(3,0.3^2)+0.5 N(-3,0.3^2)$, and $X_j$
follows the Cauchy distribution independently for all other $j$. The
counterpart for SIS for this model is to screen the predictors by
marginally performing multinomial regression. Other than SIS, only the
Kolmogorov filter and DCS are applicable to this model. Because $Y$ is
categorical, we directly take $H=Y$ for the Kolmogorov filter and apply
no further fusion. For DCS, we create a dummy variable $Y^{\mathrm
{dm}}\in\mathbb{R}^{n\times5}$ and compute the distance correlation
between $Y^{\mathrm{dm}}$ and $X_j$.
\end{longlist}

\subsection{Simulation results and conclusions}\label{sec4.2}

%
\begin{table}
\caption{Simulation results for Models 1--7. We report the minimum
number of predictors needed to keep all the useful predictors. The
numbers in the table are medians of 500 replicates. Standard errors are
in parentheses. A cell is left empty if the corresponding method is not
applied to the specific model}\label{tab1}
\tabcolsep=0pt
\begin{tabular*}{\tablewidth}{@{\extracolsep{\fill}}@{}lcccccc@{}}
\hline
&\multicolumn{3}{c}{\textbf{Model} 1} & \multicolumn{3}{c@{}}{\textbf{Model} 2}\\[-6pt]
&\multicolumn{3}{c}{\hrulefill} & \multicolumn{3}{c@{}}{\hrulefill}
\\
& \multicolumn{1}{c}{\textbf{(a)}} & \multicolumn{1}{c}{\textbf{(b)}}&\multicolumn{1}{c}{\textbf{(c)}}
&\multicolumn{1}{c}{\textbf{(a)}} &\multicolumn{1}{c}{\textbf{(b)}} &\multicolumn{1}{c@{}}{\textbf{(c)}}\\
&\multicolumn{1}{c}{$\bolds{d=2}$} & \multicolumn{1}{c}{$\bolds{d=2}$} & \multicolumn{1}{c}{$\bolds{d=2}$}
&\multicolumn{1}{c}{$\bolds{d=10}$} & \multicolumn{1}{c}{$\bolds{d=10}$} &\multicolumn{1}{c@{}}{$\bolds{d=10}$}
\\
\hline
\multicolumn{2}{@{}l}{Kolmogorov}\\
\quad $G=3$&4~(0.5)&4~(0.5)&4~(0.5)&10~(0)&10~(0)&10~(0)\\
\quad $G=4$&6~(0.9)&6~(0.9)&6~(0.9)&10~(0)&10~(0)&10~(0)\\
\quad $G=5$&12~(1.6)&12~(1.6)&12~(1.6)&10~(0)&10~(0)&10~(0)\\
\quad $G=6$&21~(3.2)&21~(3.2)&21~(3.2)&10~(0)&10~(0)&10~(0)\\
\quad Fused&2~(0.3)&2~(0.3)&2~(0.3)&10~(0)&10~(0)&10~(0)
\\[3pt]
SIS&2~(0)&1636~(93.5)&486.5~(100.5)&10~(0)&1552.5~(99.2)&1084.5~(62.9)\\
DCS&2~(0)&354~(34.8)&229~(54.0)&10~(0)&10~(0)&543~(52.8)\\
RCS&2~(0)&2~(0)&2~(0)&10~(0)&10~(0)&10~(0)\\
NIS&2~(0)&2~(0.4)&1214~(79.0)&10~(0)&10~(0)&1462.5~(92.8)\\
ELS&2~(0)&2879~(103.4)&2460.5~(87.7)&10~(0)&565~(287.8)&4401~(36.9)
\\[3pt]
QA&\\
\quad $\tau=0.5$&5~(0.6)&30.5~(5.4)&5~(0.6)&10~(0)&10~(0)&12~(0.4)\\
\quad  $\tau=0.75$&13.5~(1.9)&84.5~(13.7)&44~(7.6)&10~(0)&11~(0)&36~(2.4)
\\[6pt]
\hline
& \multicolumn{1}{c}{\textbf{Model 3}} & \multicolumn{1}{c}{\textbf{Model 4}} & \multicolumn{1}{c}{\textbf{Model 5}} & \multicolumn{1}{c}{\textbf{Model 6}} & \multicolumn{1}{c@{}}{\textbf{Model 7}}\\
& \multicolumn{1}{c}{$\bolds{d=2}$} &\multicolumn{1}{c}{$\bolds{d=3}$}& \multicolumn{1}{c}{$\bolds{d=8}$} &\multicolumn{1}{c}{$\bolds{d=2}$}&\multicolumn{1}{c@{}}{$\bolds{d=8}$}
\\
\hline
\multicolumn{2}{@{}l}{Kolmogorov}\\
\quad $G=3$&2~(0)&6~(0.8)&207.5~(27.1)&2~(0)&&\\
\quad $G=4$&2~(0)&5~(0.4)&54.5~(7.2)&&15~(0.4) \\
\quad $G=5$&2~(0)&5~(0.4)&32~(3.0)&\\
\quad $G=6$&2~(0)&7~(0.7)&25~(1.3)&\\
\quad Fused&2~(0)&3~(0)&16~(0.9)&
\\[3pt]
SIS&439.5~(38.3)&3177~(95.9)&4094~(81.0)&13~(1.7)&4661.5~(25.6)\\
DCS&260.5~(36.2)&40.5~(6.5)&22~(2.7)&1002~(89.2)&1038~(121.2)\\
RCS&2~(0)&3~(0)&3430~(124.4)\\
NIS&494~(96.4)&3258.5~(114.5)&4260.5~(55.3)\\
ELS&3247.5~(94.7)&3801~(69.1)&4510~(26.6)&3253~(96.2)
\\[3pt]
QA&\\
\quad $\tau=0.5$&50~(2.3)&17~(1.7)&1193~(129.4)\\
\quad $\tau=0.75$&70~(3.7)&1234.5~(75.4)&32.5~(1.4)\\
\hline
\end{tabular*}
\end{table}

The simulation results are reported in Table~\ref{tab1}. There are two important
conclusions.
\begin{itemize}
\item We see that the Kolmogorov filter using a single slicing works
reasonably well, and its performance is rather insensitive to the
choice of number of slices. Nevertheless, the Kolmogorov filters with
fewer slices tend to be more efficient when the underlying model is
simple, such as in Model~1 where the true model is a transformed linear
model. On the other hand, the Kolmogorov filters with more slices tend
to be more accurate when the model is complicated, such as in Model~5.
However, by combining different slicing schemes, the fused Kolmogorov
filter has the best overall performance. The fused Kolmogorov filter is
at least as good as the best $\hat K_j(G_i)$ in Models 1--3. In Models
4 and 5, where the fused Kolmogorov filter is slightly worse than the
$\hat K_j(G_i)$ with the best $G_i$, the difference is very small.
\item Compared with SIS, DCS, NIS, ELS and QA, the fused Kolmogorov
filter is either the best or one of the best, and outperforms the rest
by a large margin. This clearly shows that the fused Kolmogorov filter
is a superior screening technique.
\end{itemize}

This simulation also reveals some major drawbacks of the existing
screening methods. Although SIS, DCS, NIS and ELS work well in Models
1(a) and 2(a), variable transformation as in Models 1(b)--1(c) and
Models 2(b)--2(c) can easily destroy their performance. Models 3 and 4
are nonlinear with heavy-tailed covariates. Most screening methods
other than the fused Kolmogorov filter have too many false discoveries,
especially in Model~4.
NIS, RCS and QA are not directly applicable when we have a Poisson
regression model in Model 6. Model 5 has heteroscedasticity, which
impairs SIS, NIS, RCS and ELS.

\section{A real data example}\label{sec5}

In this section, we demonstrate the fused Kolmogorov filter on the
Tecator dataset. The Tecator dataset was collected by Tecator Infratec
Food and Feed Analyzer
working in the wavelength range 850--1050~nm by the Near Infrared
Transmission (NIT) principle. The predictors are 100 channel spectrum of
absorbances. The response is the percentage of fat in finely chopped
meat. This dataset is available at \url
{http://lib.stat.cmu.edu/datasets/tecator}. The provider of the dataset
suggested using the first 215 samples to test the performance of a
statistical method by treating 43 of them as the testing set. However,
samples \#103 and \#105 appear to be outliers, so we deleted them. Then
we standardized the response so that it has a standard deviation of 1.
We randomly chose 41 samples as our testing set in each replicate.
Also, in addition to the 100 predictors in the original dataset, we
added 4900 independent noise variables following the Cauchy distribution.

We include the fused Kolmogorov filter, DCS, SIS, QA, NIS and ELS for
comparison. First, we examine whether the screening methods can
distinguish the useful predictors from the noise variables. In the
fused Kolmogorov filter, we still consider the combination of
$G_i=3,\ldots, 6$, as in the simulation studies. For each screening
method we keep the top 100 predictors, as the ``truth'' is there are
4900 pure noise variables. We report the number of the original 100
predictors captured by screening in Table~\ref{tab2}. It is easy to see that the
fused Kolmogorov filter, DCS and NIS have much better performance in
preserving the true predictors. In particular, the fused Kolmogorov
filter has a nearly perfect screening result.

%
\begin{table}
\tabcolsep=0pt
\caption{Comparison of the screening methods on the tecator dataset.
We report the number of true predictors that are preserved after the
screening step. The numbers are averaged over 100 replicates. Standard
errors are in parentheses}\label{meat1}\label{tab2}
\begin{tabular*}{\tablewidth}{@{\extracolsep{\fill}}@{}lccccccc@{}}
\hline
& &  &  & & \multicolumn{2}{c}{\textbf{QA}} &  \\[-6pt]
& & & & & \multicolumn{2}{c}{\hrulefill} \\
&\textbf{Kolmogorov} &\textbf{DCS} &\textbf{NIS} & \textbf{SIS} &$\bolds{\alpha=0.5}$&$\bolds{\alpha=0.75}$&\textbf{ELS}\\
\hline
True predictors&99.6&75.4& 77.3&11.7&45.4&42.2&6.24\\
&(0.06)&(0.44)&(0.28)&(0.27)&(0.56)&(0.43)&(0.14)\\
\hline
\end{tabular*}
\end{table}

We further examine how variable screening helps predict the response
variable. Again, we start with the augmented dataset with the
additional 4900 pure noise variables. For a nonparametric model-free
method such as
the fused Kolmogorov filter and DCS, the prediction is made by fitting
a random forest after screening. Hence the resulting methods are called
K-RF and DCS-RF, respectively. NIS is designed based on a generalized
additive model.
So when NIS is used for variable screening, the prediction is made by a
sparse generalized additive model. We denote this method by
NIS--GroupLasso. In K-RF, DCS-RF and NIS--GroupLasso, we let $d_n=100$.

%
\begin{table}[b]
\tabcolsep=0pt
\caption{Comparison of the prediction performance on the tecator
dataset. The numbers are averaged over 100 replicates. Standard errors
are in parentheses. A paired $t$-test shows that K-RF is significantly
better than DCS-RF and NIS-RF, with $p$-values less than $1\times10^{-5}$}\label{meat2}\label{tab3}
\begin{tabular*}{\tablewidth}{@{\extracolsep{\fill}}@{}lccccc@{}}
\hline
& \textbf{K-RF} & \textbf{DCS-RF} & \textbf{NIS--GroupLasso} & \textbf{INIS--GroupLasso} & \textbf{NIS-RF}\\
\hline
Average MSE&0.097&0.102&0.195&0.187&0.103\\
&(0.009)&(0.010)&(0.019)&(0.017)&(0.010)\\
\hline
\end{tabular*}
\end{table}

Moreover, we include an iterative procedure that performs NIS and
group-lasso penalized regression repeatedly. After the initial
screening, we keep the top 100 predictors, and then we follow \citet
{FFS11} to iteratively conduct the following two-step procedure: first,
we add the predictor with the most predictive power that is not in the
selected set of predictors; second, we delete some predictors in the
selected set of predictors via group-lasso. In the deletion step, the
tuning parameter is chosen to be the largest tuning parameter that
produces an error within one standard deviation of the minimum error.
This resulting method is referred to as INIS--GroupLasso. We use the
$\tt R$ package $\tt gglasso$ [\citet{YZ14}] to fit the
group-lasso penalized additive model.

Finally, as suggested by a referee, we also include the prediction
performance for NIS followed by random forest, which is denoted by
NIS-RF. The average mean squared errors (MSE) on the testing sets are
listed in Table~\ref{tab3}. The method K-RF has significantly better performance
than all the other methods.

\section{Discussion}\label{sec6}

In this paper we have proposed the fused Kolmogorov filter and
demonstrated its superior performance over the existing screening
methods. Before concluding this work, we would like to further comment
on two main messages delivered in this paper. First, we have proposed
the slicing and fusion idea to deal with general response variables
such as continuous response variable and counts (e.g., Poisson)
response variable. In this general approach one may use a different
test statistic for testing the equivalence of two distributions to
replace the Kolmogorov--Smirnov statistic, and the resulting screening
method would be different and likely effective as well. We prefer the
Kolmogorov--Smirnov statistic because it is invariant under variable
transformation and works naturally with many different types of
covariates. Moreover, its sure screening property can be established
without assuming any special distributional property of the covariates.
Any future proposal for variable screening should possess all these
nice properties of the fused Kolmogorov filter and some nontrivial new
properties. The second message is about nonparametric screening versus
model-based screening. The vibrant research on variable screening
started with a simple model-based method, marginal correlations
screening. However, it is clear now that nonparametric model-free
variable screening should be preferred in real data analysis, unless
the user strongly believes that the data
can be fit well by a parametric model. Otherwise, nonparametric
screening methods are more robust, have wider applicability and when
combined with nonparametric learning techniques, they can provide
better prediction than a model-based method. On the other hand, an
obvious advantage of model-based screening is that its performance can
be boosted by an iterative screening and model-fitting procedure. It is
unclear how to derive a similar iterative procedure for a nonparametric
model-free screening method. It would be interesting and useful to do
so, such that we could have an iterative way to combine the fused
Kolmogorov filter or other nonparametric screening method and
nonparametric learning methods. This is an open question left for
future study. We do not expect an easy solution. Note that even for the
model-based iterative screening methods, their theoretical properties
still remain unknown.

\begin{appendix}\label{append}
\section*{Appendix: Technical proofs}
Throughout this appendix, $F$ denotes the generic cumulative
distribution function, and $f$ denotes the generic probability density
function for a random variable.

%
\begin{proposition}\label{bound1}
Consider a pair of random variables $(X,Y)$. For any interval $[a,b)$
such that $f_Y(y)>0$ for $y\in[a,b)$, we have
\begin{eqnarray*}
\inf_{y\in[a,b)} F(x\mid Y=y) &\le& F\bigl(x\mid Y\in[a,b)\bigr)\le F
\bigl(x\mid Y\in[a,b)\bigr)
\\
&\le& \sup_{y\in[a,b)} F(x\mid Y=y)
\end{eqnarray*}
for all $x$.
\end{proposition}

\begin{pf*}{Proof of Proposition~\ref{bound1}}
By definition,
\begin{eqnarray*}
F\bigl(x\mid Y\in[a,b)\bigr)&=&\frac{\int_{a}^b\int_{-\infty}^x f(x,y)\,
\diff x \,\diff y}{\int_{a}^b f_Y(y) \,\diff y}
\\
&=&\frac{\int_{a}^b\int_{-\infty}^x f(x\mid y) f_Y(y)\,\diff x\, \diff
y}{\int_{a}^b f_Y(y) \,\diff y}.
\end{eqnarray*}
Because for any $y\in[a,b)$,
\[
\inf_{y\in[a,b)} F(x\mid Y=y) \le\int_{-\infty}^x
f(x\mid y)\,\diff x\le\sup_{y\in[a,b)} F(x\mid Y=y),
\]
we have the desired conclusion.
\end{pf*}

\begin{pf*}{Proof of Lemma~\ref{Kindependent}}
We start with the first conclusion. If $X_j$ is independent of $Y$,
then $X_j$ will be independent of any $H$, which is a function of $Y$.
Therefore, $K^\bG_j=0$ for all $\bG$. Now suppose $K^{\bG}_j=0$ for
all choices of $\bG$. For any $y$, consider $H=1$ if $Y\le y$ and
$H=2$ otherwise. Because $K^{\bG}_j=0$, $X_j$ is independent of $H$.
Consequently, $\Pr(Y\le y\mid X_j)=\Pr(Y\le y)$ for all $y$, and $Y$
is independent of $X_j$.

For\vspace*{1pt} the second conclusion, suppose there exists $\bG$ such that $K^\bG
_j=0$. Then $X_j\perp H$ for the corresponding $H$. Therefore, $\Pr
(Y\le a_1\mid X_j)=\Pr(H=1\mid X_j)=\Pr(H=1)$ is a constant, which
contradicts our assumption. Therefore, we must have $K^\bG_j\ne0$.

Now we turn to the third conclusion. Because $X_j$ is not independent
of $Y$, \mbox{$K^*_j>0$}. Hence, it suffices to show that $K^\bG\rightarrow
K_j^*$ as $G\rightarrow\infty$. This is indeed true. By the
definition of $K_j^*$, for any $\varepsilon>0$, there exists
$(y_1^*,y_2^*,x^*)$ such that
\[
\bigl\llvert K_j^*-\bigl\llvert F_j\bigl(x^*\mid
y_1^*\bigr)-F_j\bigl(x^*\mid y_2^*\bigr)
\bigr\rrvert\bigr\rrvert<\varepsilon.
\]

Because $F(x^*\mid y)$ is continuous in $y$, there exists $\delta>0$
such that $\llvert F_j(x^*\mid y)-F_j(x^*\mid y_1^*)\rrvert<\varepsilon$
for any
$\llvert y-y_1^*\rrvert<\delta$. Take $\phi=\Pr(\llvert
y-y_1^*\rrvert<\delta)$. Because
$\max_{l=1,\ldots, G} \Pr(H=l)\rightarrow0$, there exists $G^*$
such that $\Pr(H=l)<\frac{\phi}{2}$ for $G>G^*$. In such cases,
there exists $[a_{l_1},b_{l_1})\subset(y_1^*-\delta,y_1^*+\delta)$.
By Proposition~\ref{bound1}, we have
\[
\bigl\llvert F_j\bigl(x^*\mid H=l_1
\bigr)-F_j\bigl(x^*\mid y_1^*\bigr)\bigr\rrvert<
\varepsilon.
\]
Similarly, for sufficiently large $G$, there exists $l_2$ such that
\[
\bigl\llvert F_j\bigl(x^*\mid H=l_2
\bigr)-F_j\bigl(x^*\mid y_2^*\bigr)\bigr\rrvert<
\varepsilon.
\]
Now note that
\[
\bigl\llvert F_j\bigl(x^*\mid H=l_1
\bigr)-F_j\bigl(x^*\mid H=l_2\bigr)\bigr\rrvert\le
K^\bG_j\le K_j^*.
\]

Hence
\begin{eqnarray*}
&&\bigl\llvert K_j^*-K^\bG_j\bigr\rrvert
\\
&&\qquad \le \bigl\llvert F_j\bigl(x^*\mid y_1^*
\bigr)-F_j\bigl(x^*\mid y_2^*\bigr)\bigr\rrvert+
\varepsilon- \bigl\llvert F_j\bigl(x^*\mid H=l_1
\bigr)-F_j\bigl(x^*\mid H=l_2\bigr)\bigr\rrvert
\\
&&\qquad \le\sum_{i=1,2}\bigl\llvert F_j
\bigl(x^*\mid y_i^*\bigr)-F_j\bigl(x^*\mid
H=l_i\bigr)\bigr\rrvert+\varepsilon
\\
&&\qquad <3\varepsilon.
\end{eqnarray*}
Therefore, the conclusion follows.
\end{pf*}

\begin{pf*}{Proof of Lemma~\ref{Knormal}}
Because $K^*_j$ and $K^{\bG}_j$ are invariant under monotone
transformations, it suffices to consider the case $g_1(t)=t$,
$g_2(t)=t$, and hence $X_j$ and $Y$ are jointly normal. Let $f_y(y)$ be
the probability density function of~$Y$, which is standard normal. For
the first conclusion, note that if $\rho_j=0$, then $X_j$ is
independent of $Y$ and $K^*_j=0$. On the other hand, if $\rho_j\ne0$,
$X_j\mid Y=y\sim N(\rho_j y, (1-\rho_j^2))$. Therefore, $F_j(x\mid
y)=\Phi(\frac{x-\rho_j y}{\sqrt{1-\rho_j^2}})$. It follows that
$K^*_j\ge\lim_{y\rightarrow-\infty} F_j(0\mid y)- \lim_{y\rightarrow
\infty} F_j(0\mid y)=1$. Meanwhile, by definition,
$K^*_j\le1$. Therefore, $K^*_j=1$.

For the second conclusion, again by $X_j\mid Y\sim N(\rho_j Y, (1-\rho
_j^2))$ and $F_j(x\mid y)=\Phi(\frac{x-\rho_j y}{\sqrt{1-\rho
_j^2}})$, we have
\begin{eqnarray*}
F_j(x\mid H=l)&=&\frac{\Pr(X_j\le x,H=l)}{\Pr(H=l)}
\\
&=&G\int_{a_{l-1}}^{a_l} \Phi\biggl(\frac{x-\rho_j y}{\sqrt{1-\rho
_j^2}}
\biggr)f(y) \,\diff y
\\
&\in&\biggl[\Phi\biggl(\frac{x-\rho_j a_{l-1}}{\sqrt{1-\rho_j^2}}\biggr
),\Phi\biggl(\frac{x-\rho_j a_l}{\sqrt{1-\rho_j^2}}
\biggr)\biggr].
\end{eqnarray*}

Now, for $1\le l<m\le G$,
\begin{eqnarray*}
&&\sup_{x}\bigl\llvert F_j(x\mid
H=l)-F_j(x\mid H=m)\bigr\rrvert
\\
&&\qquad \le \sup_x \biggl(\Phi\biggl(\frac{x-\rho_j a_{l-1}}{\sqrt{1-\rho
_j^2}}\biggr)-
\Phi\biggl(\frac{x-\rho_j a_{m}}{\sqrt{1-\rho_j^2}}\biggr)\biggr)
\\
&&\qquad =2\Phi\biggl(\frac{\rho_j (a_m-a_{l-1})}{\sqrt{1-\rho_j^2}}\biggr)-1.
\end{eqnarray*}
On the other hand,
\begin{eqnarray*}
&&\sup_{x}\bigl\llvert F_j(x\mid
H=1)-F_j(x\mid H=G)\bigr\rrvert
\\
&&\qquad \ge \sup_x \biggl(\Phi\biggl(\frac{x-\rho_j a_{1}}{\sqrt{1-\rho
_j^2}}\biggr)-
\Phi\biggl(\frac{x-\rho_j a_{G-1}}{\sqrt{1-\rho_j^2}}\biggr)\biggr)
\\
&&\qquad =2\Phi\biggl(\frac{\rho_j (a_1-a_{G-1})}{\sqrt{1-\rho_j^2}}\biggr
)-1\ge2\Phi\biggl(\frac{\rho_j (a_m-a_{l-1})}{\sqrt{1-\rho_j^2}}
\biggr)-1
\\
&&\qquad \ge\sup_{x}\bigl\llvert F_j(x\mid
H=l)-F_j(x\mid H=m)\bigr\rrvert.
\end{eqnarray*}
Therefore,
\begin{eqnarray*}
K^\bG_j &=& \sup_{x}\bigl\llvert
F_j(x\mid H=1)-F_j(x\mid H=G)\bigr\rrvert.
\end{eqnarray*}
Moreover, note that $a_1=-a_{G-1}$. By checking the derivatives, we have
\begin{eqnarray*}
K^\bG_j &=& \bigl\llvert F_j(0\mid
H=1)-F_j(0\mid H=G)\bigr\rrvert.
\end{eqnarray*}
Hence
\begin{eqnarray*}
K^\bG_j&=&G\biggl(\int_{-\infty}^{a_1}
\Phi\biggl(\frac{-\rho_j y}{\sqrt
{1-\rho_j^2}}\biggr)f(y)\,\diff y-\int_{a_{G-1}}^{\infty}
\Phi\biggl(\frac{-\rho
_j y}{\sqrt{1-\rho_j^2}}\biggr)f(y)\,\diff y\biggr)
\\
&=&G\biggl(\int_{-\infty}^{a_1} \Phi\biggl(
\frac{-\rho_j y}{\sqrt{1-\rho
_j^2}}\biggr)f(y)\,\diff y-\int_{-\infty}^{a_1}
\biggl(1-\Phi\biggl(\frac{-\rho_j
y}{\sqrt{1-\rho_j^2}}\biggr)\biggr)f(y)\,\diff y\biggr)
\\
&=&G\biggl(\int_{-\infty}^{a_1} \biggl(2\Phi\biggl(
\frac{-\rho_j y}{\sqrt{1-\rho
_j^2}}\biggr)-1\biggr)f(y)\,\diff y\biggr).
\end{eqnarray*}
Because $a_1\le0$, $\Phi(\frac{-\rho_j y}{\sqrt{1-\rho_j^2}})$
is strictly increasing in $\rho_j$ for each $y\in(-\infty,a_1)$.
Hence $K^\bG_j$ is strictly increasing in $\rho_j$.
\end{pf*}

Now we prove Theorem~\ref{KH1}. In order to prove this theorem, we
need the following lemmas.

%
\begin{lemma}\label{quantile-bound}
If $\hat a_l$ is the sample $\frac{l}{G}$th quantile for $Y$, then
with a probability greater than $1-C\exp(-C\frac{n}{G^2})$, we have
%
%
\begin{equation}
\label{quantile} \Pr(\hat a_l\le Y<\hat a_{l+1})<
\frac{2}{G}.
\end{equation}
\end{lemma}

%
\begin{lemma}\label{lemmaKO}
Under the conditions in Theorem~\ref{KH1}, for any $\varepsilon>0$, we have:
\begin{longlist}[(2)]
\item[(1)]
%
%
\begin{eqnarray}\label{KO}
&& \Pr\bigl(\bigl\llvert\hat K^{(o)}_j-K^{(o)}_j
\bigr\rrvert\ge N\varepsilon\bigr)
\nonumber\\[-8pt]\\[-8pt]\nonumber
&&\qquad \le CN\bigl(\log^2{n}\bigr)\exp
\biggl(-C\frac{n\varepsilon^2}{\log{n}}\biggr)
+CN\bigl(\log^2{n}\bigr)\exp
\biggl(-C\frac
{n}{\log^2{n}}\biggr);
\end{eqnarray}
\item[(2)]
%
%
\begin{equation}
\label{K} \Pr\bigl(\llvert\hat K_j-K_j\rrvert\ge
N\varepsilon\bigr)\le CN\bigl(\log^2{n}\bigr)\exp\biggl(-C
\frac
{n\varepsilon^2}{\log{n}}\biggr).
\end{equation}
\end{longlist}
\end{lemma}

%
\begin{lemma}\label{C2}
Under the conditions in Theorem~\ref{KH1}, we have
\[
\Pr\bigl(\bigl\llvert K_j-K^{(o)}_j\bigr
\rrvert\ge N\Delta_\bS/4\bigr)\le CN\exp\biggl(-C\frac{n}{\log^2{n}}
\biggr).
\]
\end{lemma}

With Lemmas~\ref{quantile-bound}--\ref{C2}, we are ready to prove
Theorem~\ref{KH1}.

\begin{pf*}{Proof of Theorem~\ref{KH1}}
We\vspace*{1pt} first consider the first conclusion. Note that if $\llvert \hat
K^{(o)}_j-K^{(o)}_j\rrvert<N\Delta_\bS/4$ for all $j$, we must have
$\bD
\subset\hat\bD$. This is indeed true because, combining it with
condition (C1), we have
\begin{eqnarray*}
\hat K^{(o)}_j&>& K^{(o)}_j-N
\Delta_\bS/4\ge\max_{j\notin\bS
}K^{(o)}_j+N
\Delta/4\qquad\mbox{for $j\in\bS$},
\\
\hat K^{(o)}_j&<& K^{(o)}_j+N
\Delta_\bS/4\le\max_{j\notin\bS
}K^{(o)}_j+N
\Delta/4\qquad\mbox{for $j\notin\bS$}.
\end{eqnarray*}
Hence, $\bS\subset\hat\bD$ and $\bD\subset\hat\bD$.

By (\ref{KO}), we have the desired conclusion.

For the second conclusion, we again have that, if $\llvert \hat
K_j-K^{(o)}_j\rrvert<N\Delta_\bS/4$ for all~$j$, we must have $\bD
\subset
\hat\bD$.

Combining (\ref{K}) and Lemma~\ref{C2}, we have
\begin{eqnarray*}
&& \Pr\bigl(\bigl\llvert\hat K_j-K^{(o)}_j\bigr
\rrvert>N\Delta_\bS/4\bigr)
\\
&&\qquad \le CN\exp\biggl(-C\frac{n}{\log
^2{n}}
\biggr)+CN\bigl(\log^2{n}\bigr)\exp\biggl(-C\frac{n\Delta_\bS^2}{\log
{n}}\biggr).
\end{eqnarray*}
Then we have the desired conclusion.
\end{pf*}

\begin{pf*}{Proof of Lemma~\ref{quantile-bound}}
First, we show that, under the event $A=\break \sup_y \llvert \hat
F_y(y)-F_y(y)\rrvert\le
\frac{1}{8G}$, we must have (\ref{quantile}). Indeed, under event $A$,
\begin{eqnarray*}
&&\Pr(\hat a_l\le Y<\hat a_{l+1})
\\
&&\qquad =\Pr\biggl(
\frac{l}{G}\le\hat F_y(Y)<\frac{l+1}{G}\biggr)
\\
&&\qquad \le \Pr\biggl(\frac{l}{G}-\frac{1}{8G}\le F_y(Y)<
\frac{l+1}{G}+\frac
{1}{8G}\biggr)=\frac{5}{4G}<
\frac{2}{G}.
\end{eqnarray*}
Then note $\Pr(A)\ge1-C\exp(-C\frac{n}{G^2})$ by the
Dvoretzky--Kiefer--Wolfowitz inequality, and the conclusion follows.
\end{pf*}

\begin{pf*}{Proof of Lemma~\ref{lemmaKO}}
We first show (\ref{KO}). Consider a single partition $\bG_i$ with
$G_i$ intervals bounded by the theoretical quantiles. Then
$H_i^{(o)}=g$ if and only if $Y$ is between its $\frac{g}{G_i}$th and
$\frac{g+1}{G_i}$th quantile. Set $K^{(o)}(G_i;g,g')=\sup_x
\llvert F_j(x\mid H^{(o)}_i=g)-F_j(x\mid H^{(o)}_i=g')\rrvert$. Then
$\Pr
(H^{(o)}_i=g)=\Pr(H^{(o)}_i=g')=\frac{1}{G_i}$. By Lemma A1 in
\citet{Kfilter}, we have
\begin{eqnarray*}
&& \Pr\bigl(\bigl\llvert\hat K_j^{(o)}\bigl(G_i;g,g'
\bigr)-K_j^{(o)}\bigl(G_i;g,g'
\bigr)\bigr\rrvert\ge\varepsilon\bigr)
\\
&&\qquad \le C\exp\biggl(-Cn\frac{\varepsilon
^2}{G_i}\biggr)+C
\exp\biggl(-C\frac{n}{G_{i}^2}\biggr).
\end{eqnarray*}
Then if $\llvert \hat K_j^{(o)}(G_i;g,g')-K_j^{(o)}(G_i;g,g')\rrvert
\le\varepsilon$
for all $g,g'$, we must have
\begin{eqnarray*}
\bigl\llvert\hat K_j^{(o)}-K_j^{(o)}
\bigr\rrvert&=&\Bigl\llvert\max_{g,g'} \hat
K_j^{(o)}\bigl(G_i;g,g'\bigr)-
\max_{g,g'} K_j^{(o)}\bigl(G_i;g,g'
\bigr)\Bigr\rrvert
\\
&\le& \max_{g,g'}\bigl\llvert\hat K_j^{(o)}
\bigl(G_i;g,g'\bigr)-K_j^{(o)}
\bigl(G_i;g,g'\bigr)\bigr\rrvert\le\varepsilon.
\end{eqnarray*}
Therefore,
\begin{eqnarray*}
&& \Pr\bigl(\bigl\llvert\hat K_j^{(o)}(G_i)-K_j^{(o)}(G_i)
\bigr\rrvert>\varepsilon\bigr)
\\
&&\qquad \le  CG_i^2\exp\biggl(-Cn\frac{\varepsilon^2}{G_i}
\biggr)+CG_i^2\exp\biggl(-C\frac
{n}{G_{i}^2}\biggr)
\\
&&\qquad \le C\bigl(\log^2{n}\bigr)\exp\biggl(-Cn\frac{\varepsilon^2}{\log{n}}
\biggr)+C\bigl(\log^2{n}\bigr)\exp\biggl(-C\frac{n}{\log^2{n}}\biggr).
\end{eqnarray*}
Finally, note that
\begin{eqnarray*}
\Pr\bigl(\bigl\llvert\hat K_j^{(o)}(G_i)-K_j^{(o)}(G_i)
\bigr\rrvert> N\varepsilon\bigr) &\le&\sum_i \Pr\bigl(
\bigl\llvert\hat K_j^{(o)}(G_i)-K_j^{(o)}(G_i)
\bigr\rrvert>\varepsilon\bigr),
\end{eqnarray*}
and the conclusion follows. For (\ref{K}), redefine $H_i=l$ if $Y$ is
with in the $\frac{l}{G_i}$th and $\frac{l+1}{G_i}$th sample
quantiles. Note that
\begin{eqnarray*}
&&\Pr\bigl(\bigl\llvert\hat K_j\bigl(G_i;g,g'
\bigr)-\hat K_j\bigl(G_i;g,g'\bigr)\bigr
\rrvert\ge\varepsilon\bigr)
\\
&&\qquad \le  \sum_{l=g,g'}\Pr\Bigl(\sup_x
\bigl\llvert\hat F_j(x\mid H_i=l)-F_j(x\mid
H_i=l)\bigr\rrvert\ge\varepsilon/2\Bigr)
\\
&&\qquad  \le C\exp\biggl(-Cn\frac{\varepsilon^2}{G_i}\biggr),
\end{eqnarray*}
where the last inequality follows from the Dvoretzky--Kiefer--Wolfowitz
inequality and the fact that there are $\frac{n}{G_i}$ observations in
the $g$th and $g'$th slice, respectively. Then because $G_i\le\lceil
\log{n}\rceil$, we have the desired conclusion. Finally, (\ref{K})
can be proven in a similar way to (\ref{KO}).
\end{pf*}

\begin{pf*}{Proof of Lemma~\ref{C2}}
First, note that
%
%
\begin{equation}
\label{eq1} \Pr\bigl(\bigl\llvert K_j-K^{(o)}_j
\bigr\rrvert\ge N\Delta_\bS/4\bigr)\le\sum
_i \Pr\bigl(\bigl\llvert K_j(G_i)-K^{(o)}_j(G_i)
\bigr\rrvert\ge\Delta_\bS/4\bigr).
\end{equation}
Therefore, we establish a bound for $\Pr(\llvert
K_j(G_i)-K^{(o)}_j(G_i)\rrvert\ge
\Delta_\bS/4)$.

Define
\[
K_{0j}=\sup_x\Bigl(\sup_y
F(x\mid y)-\inf_y F(x\mid y)\Bigr).
\]
For any $x$ and $l$, we have
\[
\inf_y F_j(x\mid y)\le F_j(x\mid
H=l)\le\sup_y F_j(x\mid y).
\]
It follows that $K_j(G_i)\le K_{0j}$ and $K_j^{(o)}(G_i)\le K_{0j}$.
Moreover, for any $\varepsilon>0$, there exists $(x^*,y_1^*,y_2^*)$ such that
\[
K_{0j}\le F_j\bigl(x^*\mid y_1^*
\bigr)-F_j\bigl(x^*\mid y_2^*\bigr)+\varepsilon.
\]

Then there exists $[a_{l_i},a_{l_i+1})\in\bG$ such that $y_i^*\in
[a_{l_i},a_{l_i+1})$. Hence,
\[
K_{0j}-K^{(o)}_j(G_i)\le\varepsilon+
\sum_{i=1,2}\bigl\llvert F_j\bigl(x^*\mid
y_1^*\bigr)-F_j\bigl(x^*\mid H=l_i\bigr)
\bigr\rrvert\le\varepsilon+\Delta_{\bS}/8,
\]
where the last inequality follows from condition (C2) and
Proposition~\ref{bound1}. Because $\varepsilon$ is arbitrary, we have
$K_{0j}-K^{(o)}_j(G_i)\le\Delta_{\bS}/8$ and hence $K_j\le
K^{(o)}_j(G_i)+\Delta_{\bS}/8$. On the other hand, suppose
\[
K^{(o)}_j(G_i)=F_j
\bigl(x_0\mid H^{(o)}_i=l_1
\bigr)-F_j\bigl(x_0\mid H^{(o)}_i=l_2
\bigr).
\]
Set\vspace*{-1pt} $y_1^*$ such that $y_1^*\in\{y\dvtx  H^{(o)}_i=l_1\}$ and $\inf_{y\dvtx
H^{(o)}_i=l_1}F_j(x\mid y)=F_j(x\mid y_1^*)$. Note that $y_1^*$ can be
$+\infty$ or $-\infty$. Then there exists $l_1'$ such that $y_1^*\in
\{H^{(o)}_i=l_1\}\cap\{H_i=l_1'\}$. Also define $y_2^*$ as the number
that $y_2^*\in\{y\dvtx  H^{(o)}_i=l_2\}$ and $\sup_{y\dvtx
H^{(o)}_i=l_1}F_j(x\mid y)=F_j(x\mid y_2^*)$. Note\vspace*{1pt} that $y_2^*$ can be
$+\infty$ or $-\infty$ as well. Then there exists $l_2'$ such that
$y_2^*\in\{H^{(o)}_i=l_2\}\cap\{H_i=l_2'\}$.

We claim that if $\Pr(H_i=l_k')\le2/G$, we must have $K_j(G_i)\ge
K^{(o)}_j(G_i)-\Delta_\bS/4$. Indeed, by Proposition~\ref{bound1},
\[
K_j\ge\inf_{y\dvtx  H_i=l_1'} F_j(x_0
\mid y)-\sup_{y\dvtx  H_i=l_2'} F_j(x_0\mid y).
\]
Then by condition (C2), if $\Pr(H_i=l_k')\le2/G$, we must have
\begin{eqnarray*}
K_j &\ge&\inf_{y\dvtx  H_i=l_1'} F_j(x_0
\mid y)-\sup_{y\dvtx  H_i=l_2'} F_j(x_0\mid y)
\\
&\ge& F_j\bigl(x_0\mid y_1^*
\bigr)-F_j\bigl(x_0\mid y_2^*\bigr)
\\
&\ge& \inf_{y\dvtx  H^{(o)}_i=l_1} F_j\bigl(x^*\mid y\bigr)-
\Delta_\bS/8-\sup_{y\dvtx
H^{(o)}_i=l_2} F_j\bigl(x^*
\mid y\bigr)-\Delta_\bS/8
\\
&\ge& K_j^{(o)}-\Delta_\bS/4,
\end{eqnarray*}
where the last inequality again follows from condition (C2) and
Proposition~\ref{bound1}.

By Lemma~\ref{quantile-bound}, we have
\[
\Pr\bigl(\Pr\bigl(H_i=l_k'\bigr)> 2/G
\bigr)\le C\exp\biggl(-C\frac{n}{G_i^2}\biggr).
\]
Therefore,
%
%
\begin{eqnarray}\label{eq2}
\Pr\bigl(\bigl\llvert K_j(G_i)-K^{(o)}_j(G_i)
\bigr\rrvert\ge\Delta_{\bS}/4\bigr) &\le& C\exp\biggl(-C\frac
{n}{G_i^2}
\biggr)
\nonumber\\[-8pt]\\[-8pt]\nonumber
&\le& C\exp\biggl(-C\frac{n}{\log^2{n}}\biggr).
\end{eqnarray}

Combining (\ref{eq1}) and (\ref{eq2}) we have the desired conclusion.
\end{pf*}

\begin{pf*}{Proof of Lemma~\ref{correlated}}
For the first conclusion, note that
\[
\pmatrix{ T_y(Y)
\cr
\bT(\bX)} \sim N \left(\pmatrix{ 0
\cr
0 },
\pmatrix{ \bbeta^\T\bSigma\bbeta+\sigma^2&
\bbeta^\T\bSigma
\cr
\bSigma\bbeta&\bSigma} \right).
\]

Straightforward calculation shows that
\[
\bigl\llvert\cor\bigl(T_y(Y),T_j(X_j)
\bigr)\bigr\rrvert=\frac{\llvert \alpha_j\rrvert}{\sqrt{\bbeta^\T
\bSigma
\bbeta+\sigma^2}}
\]
is monotone in $\llvert \alpha_j\rrvert$.
Now that, for any $G_i$, $K^{(o)}_j(G_i)$ is invariant under strictly
monotone transformations. Therefore, by the second conclusion in
Lemma~\ref{Knormal}, $K^{(o)}_j(G_i)$ is strictly increasing in
$\llvert \alpha_j\rrvert$, and the conclusion follows.

For the second conclusion, note that when $\bSigma$ is blockwise
independent, we must have $\balpha_{\bD^C}=0$.

For the third conclusion, note that for $j>d$, we have $\alpha_j=\rho
^{j-d}\alpha_d$. When $j>d+\frac{\log{\min_{j\in\bD}\llvert \alpha
_j\rrvert/\llvert \alpha_d\rrvert}}{\log{\llvert \rho\rrvert}}$,
we must have $\llvert \alpha_j\rrvert<\min_{j\in\bD}\llvert \alpha
_j\rrvert$, and the conclusion follows.

For the third conclusion, write $\bSigma=(1-\rho)\bI+\rho\bJ$,
where $\bJ$ is a $p\times p$ matrix of 1. Then $\bSigma^{-1}=(1-\rho
)^{-1}I-\rho[\{1+(p-1)\rho\}(1-\rho)]^{-1}\bJ$. Write $c=\mathrm
{1}^\T\bbeta=\sum_{j \in\bS} \beta_j$.
For any $j \in\bS$, we have
{$\beta_j=- \rho[\{1+(p-1)\rho\}(1-\rho)]^{-1}c$}. Thus $\bD
\subseteq\bS\Leftrightarrow\mathrm{1}^\T\bbeta=0$.

For the fourth conclusion, note that for any $j\in\bD$, we have
$\alpha_j=(1-\rho)\beta_j+\rho\mathrm{1}^\T\bbeta$, while for
$j\notin\bD$, we have $\alpha_j=\rho\mathrm{1}^\T\bbeta$. Hence,
when $\rho>0$ and $\beta_j$ has the same sign for all $j\in\bD$, we
have $\Delta_{\bD}>0$.
\end{pf*}
\end{appendix}

\section*{Acknowledgments}
We are grateful to the Editor, the Associate Editor and two referees
for helpful suggestions. We thank Professor Lan Wang for providing the
implementation of QA and Professor Yichao Wu for providing the
implementation of ELS.


%

\printaddresses
\end{document}